\documentclass[aps,pra,showpacs,amsmath,amssymb,twocolumn,10pt]{revtex4-1}
\usepackage{epsfig,graphicx,color}
\begin{document}
\newfont{\Yfont}{cmti10 scaled 2074}
\newcommand{\Y}{\hbox{{\Yfont y}\phantom.}}

\title{Effective range from tetramer dissociation data for cesium atoms}
\author{M. R. Hadizadeh$^1$}
% \email{hadizade@ift.unesp.br}
\author{M. T. Yamashita$^1$}
% \email{yamashita@ift.unesp.br}
\author{Lauro Tomio$^{1,2}$}
% \email{tomio@ift.unesp.br}
\author{A. Delfino$^3$}
% \email{delfino@if.uff.br}
\author{T. Frederico$^4$}
% \email{tobias@ita.br}

\affiliation{
$^1$Instituto de F\'{\i}sica Te\'orica,
Universidade Estadual Paulista, 01140-070, S\~ao Paulo, SP, Brazil\\
$^2$Universidade Federal do ABC, 09210-170, Santo Andr\'e, SP, Brazil \\
$^3$Instituto de F\'{\i}sica, Universidade Federal Fluminense, 24210-346,
Niter\'oi, RJ, Brazil\\
$^4$Instituto Tecnol\'ogico de Aeron\'autica, DCTA, 12228-900, S\~ao
Jos\'e dos Campos, SP, Brazil
}

\date{\today}

\begin{abstract}
The shifts in the four-body recombination peaks, due to an
{effective range correction to the zero-range model close to the
unitary limit}, are obtained and used to extract the corresponding
effective range of a given atomic system. 
The approach is applied to an ultracold gas of cesium atoms close to 
broad Feshbach resonances, where deviations of experimental
values from universal model predictions are associated to 
effective range corrections. 
The effective range correction is extracted, with a weighted average
given by 3.9$\pm 0.8 R_{\mbox{vdW}}$, where $R_{\mbox{vdW}}$ is the
van der Waals length scale; which is consistent with the van der Waals 
potential tail for the $Cs_2$ system. 
The method can be generally applied to other cold atom experimental 
setups to determine the contribution of the effective range to the 
tetramer dissociation position.
\end{abstract}
 \maketitle

\section{Introduction}
The fundamental quantum properties of atomic systems are under intense investigation
in recent cold-atom experiments, in which long-time predicted universal aspects of
few-body physics can be probed. In this regard, the increasing number of three-body
bound states (trimers), emerging when the absolute value of the two-body scattering
length is moved to infinity, known as Efimov effect~\cite{EfiPLB70}, was confirmed in
different cold-atom laboratories~\cite{grimm,2009knoop,Zaccanti,2009Barontini}.

Actually, by extending  to more than three particles the search for fundamental quantum
behavior, in the realm of four-body physics it is of interest to test the positions of
four-atom resonant recombination peaks~\cite{ferlaino2009,pollack,BerPRL11,FerFBS11},
which appear as the two-body scattering length is tuned close to an $s-$wave Feshbach
resonance (see e.g. \cite{ChinRMP10}).

On the theoretical side, the model independence of tetramer properties, when few
physical scales are fixed, was recently numerically established by our
group~\cite{hadi2011,hadi2012}, through the analysis of observed correlations between
tetramer observables, by considering a renormalized zero-range (RZR) approach. These
correlations were realized through appropriate scaling functions, from where one can
verify that the RZR results are  consistent with several other model
calculations~\cite{stechernature,HamPRA2007,PlaPRA2009,DelPRA12}.

Despite the success of the model independence of the tetramer scaling functions, the
experimental data for cesium atoms~\cite{ferlaino2009,BerPRL11,FerFBS11} are shifted
with respect to the correlation function relating the peaks of successive tetramer
resonances for a zero-range force~\cite{hadi2011}. Among the possible effects to
explain such a deviation from the results obtained with zero-range two-body interaction,
it is natural to think in a way to introduce the correction due to a finite effective range in
the model, as suggested in Ref.~\cite{freFBS12}.

The experimental data for three-body cold-atom reactions show that
Efimov resonances and recombination minima survive in the
intermediate regime when the scattering length $a$ exceeds the
characteristic interaction range by a relatively small
factor~\cite{FerPHYS2010}. But the positions are shifted away from
the naive predictions  based on the universal
theory~\cite{HamPRA2007,PlaPRA2009} and finite range potentials have
been used \cite{ThoPRA2008,JonPRL2010} to address such issues. On
the other hand, the quantitative reproduction of the experimental
findings for the positions of the peaks of the tetramers resonances
has been not yet addressed in detail.

Within our aim in clarifying the shift in the experimental data with
respect to the zero-range predictions for the positions of the
tetramer resonances at the dissociation point,
we present an approach on how to introduce the effective range
correction in lowest order in the zero-range four-body theory. We
obtain the correction to the correlation of the positions of two
successive resonant four-boson recombination peaks. The associated
scaling function is not fully determined by the trimer properties
and requires a four-boson scale~\cite{hadi2011}. Furthermore, we are
showing by our calculations that the effective range
correction is not completely parameterized by the trimer and
tetramer short-range scales, as one could naively expect.  The
contribution of a non-zero effective range affects the observed
scaling plots, by moving the corresponding positions in relation to
the zero-ranged ones (see e.g. Refs.~\cite{ThoPRA2008,FrePRA99} in
the case of trimers). Here, it will be shown quantitatively how the
effective range affects the scaling between negative scattering
length ratios, corresponding to the dissociation positions of a
trimer and related tetramers (for two nearby tetramer states,
associated with a given trimer state).

Relevant to the understanding of the  effective range
contribution to the resonance positions, we should recall the nice
and clear discussion by Chin et al.~\cite{ChinRMP10} on the role of
broad (entrance-channel dominated) and narrow (closed-channel
dominated) Feshbach resonances. As discussed, a Feshbach resonance
strongly dominated by the entrance channel allows a description of a
two-atom system in terms of a single-channel short-ranged model with
a van der Waals tail. The broad $s-$wave Feshbach resonance of
cesium~\cite{ferlaino2009,BerPRL11,FerFBS11}, where tetramer
dissociations at the four-body continuum threshold were observed, is
dominated by an entrance-channel. In this case, when the strength
parameter is very large ($s_{res}>>1$), the single-channel
description is applicable~\cite{ChinRMP10} and consequently, the
effective range should be found around the values obtained for van
der Waals like potentials, tuned to the scattering length of the
resonance. Indeed, as we are going to detail in the present work,
the estimated values we have obtained for the effective ranges, from
the positions of tetramer resonances, are somewhat close to the
corresponding effective range of that class of potentials.
 Within our approach to extract the effective range, we have
considered broad Feshbach resonances appearing in ultracold gas of
cesium atoms, leading to positive effective ranges. In contrast,
narrow resonances, as discussed by Petrov~\cite{PetPRL04}, should
lead to negative effective ranges. In the case of systems with
narrow resonances and denser spectrum, we should also mention the
recent work by S{\o}rensen et al.~\cite{SorPRA12}.

The present work is organized as follows. In Sect. \ref{sec:2}, the
formalism for two, three and four-boson systems is presented, where
we adressed the effective range expansion of the two-body scattering
amplitude, the  subtracted Skorniakov and Ter-Martirosian (STM)
equation and  Faddeev-Yakubovsky (FY) equations. In Sect. \ref{sec:3}
the shifts in the position of four-body resonant loss peaks from the
solution of the STM and FY equations with the effective range
correction are given. In Sect. \ref{sec:4}, we performed an analysis of 
the positions of the four-boson resonance peaks for the cesium data 
close to wide Feshbach resonances. The effective range, shown to be 
compatible with experiments, are compared to the one obtained from 
van der Waals like potentials for the $Cs_2$ system.
In Sect. \ref{concl} we present our conclusions and perspectives.

\section{Formalism for two, three and four-boson systems\label{sec:2}}
\subsection{Two-body scattering amplitude with effective range correction}

The $s-$wave two-boson scattering amplitude input to the subtracted
forms of the trimer Skorniakov and Ter-Martirosian
equation~\cite{AdhPRL95} and the tetramer coupled Faddeev-Yakubovsky
equations~\cite{YamEPL06} for negative energies $\epsilon$, with the
on-shell momentum given by $k={\rm i}\sqrt{-\epsilon}$, can be
written as
\begin{eqnarray} \tau(\epsilon) \approx
{\frac{1/(2\pi^2)}{-\sqrt{-\epsilon}\pm\sqrt{-\epsilon_2}}}
\left[1+{\frac{r_0}{2}}\left(\sqrt{-\epsilon}\pm\sqrt{-\epsilon_2}\right)\right] ,
\label{texp}
\end{eqnarray}
in the lowest order of the effective range $r_0$.
In the above, the pole of the scattering amplitude is fixed by the bound
$(+)$ or virtual state $(-)$ energy, $k^2=\epsilon=-|\epsilon_2|$. The inverse of the
scattering length, up to the same order, is
\begin{eqnarray}
\frac{1}{a}= \pm\sqrt{|\epsilon_2|}- \frac{r_0}{2} |\epsilon_2| \ .
\label{a}
\end{eqnarray}
The two-body amplitude (\ref{texp}) provides a correction to the STM
and FY equations in leading order in $r_0$, and as we verified the
subtracted forms of these equations  are able to provide finite
results dealing with the extra power of momentum in the amplitude.

In the following, we present some details on the formalism for the
subtracted equations we have used, together with our approach to
 find the effective range correction for the position of the
tetramer resonances.

\subsection{Subtracted STM equation}

The Faddeev components for three-boson bound states for the
zero-range potential can be rewritten in terms of spectator
functions, which are solutions of a subtracted Skorniakov and
Ter-Martirosian (STM) integral equation, given by\cite{AdhPRL95}:
\begin{eqnarray}
|\, {\cal K}_{ij,k} \,\rangle = 2\,\tau(\epsilon_{ij,k})\,  {\cal
G}^{(3)}_{ij;ik} \,\, |\, {\cal K}_{ik,j} \,\rangle , \label{FE1}
\end{eqnarray}
where $\epsilon_{ij,k}$ is the $(ij)$ subsystem energy in the
three-body system. Note that we are going to introduce the two-boson
amplitude (\ref{texp}) carrying the effective range correction in
(\ref{FE1}).  The subtraction of the free 3-body resolvent regulates
STM equation, and projected operator, ${\cal G}^{(3)}$, is:
\begin{eqnarray}
{\cal G}^{(3)}_{ij;ik}=\langle\chi_{ij}|
\left(\left[E-H_0\right]^{-1}-\left[-\mu_{3}^2-
H_0\right]^{-1}\right)|\chi_{ik}\rangle, \label{G3}
\end{eqnarray}
where $H_0$ is the three-body free hamiltonian and  $-\mu^2_3$ is
the subtraction energy scale. The form factor appearing in the
two-body T-matrix of the pair $(ij)$, in the relative momentum
$\textbf{p}_{ij}$ id given by $\langle
\textbf{p}_{ij}|\chi_{ij}\rangle=1$. The three-body regularization
parameter $\mu_3$ keeps under control the Thomas-collapse, and
determine the three-boson observables, and can be parameterized by
one $s-$wave physical quantity(see e.g. \cite{YamPRA02}). The
subtracted STM equations was also used to calculate the position of
triatomic continuum resonances for large negative scattering lengths
\cite{BriPRA04}, which were observed by the Innsbruck group
\cite{grimm}, through three-atom recombination peak in a cold cesium
gas close to a Feshbach resonance.

\subsection{ Subtracted FY equations}

The tetramer energies are found by solving the Faddeev-Yakubovsky
(FY) equations properly regulated in the limit of the zero-range
interaction, which are written as a set of coupled subtracted
integral equations \cite{YamEPL06,hadi2011,hadi2012} given by:
\begin{eqnarray}
|\, {\cal K}_{ij,k}^{\,\,\, l} \,\rangle &=&
2\,\tau(\epsilon_{ij,k}^{\,\,\, l})\,\, \Biggl[ {\cal
G}^{(3)}_{ij;ik} \,\, |\, {\cal K}_{ik,j}^{\,\,\, l} \,\rangle + \nonumber \\
&& +{\cal G}^{(4)}_{ij;ik} \,\, \biggl( |\, {\cal K}_{ik,l}^{\,\,\,
j} \,\rangle + |\, {\cal H}_{ik, jl} \,\rangle \biggr) \Biggr] ,
\label{FYE1} \\ |\, {\cal H}_{ij, kl} \,\rangle &=&
\tau(\epsilon_{ij,kl}) \,\, {\cal G}^{(4)}_{ij;kl}
 \,\, \Biggl[
2\,\, |\, {\cal K}_{kl,i}^{\,\,\, j} \,\rangle+ |\, {\cal H}_{kl,
ij} \,\rangle \Biggr]. \label{FYE2}
\end{eqnarray}
The subtractions in the kernel come at the level of the free 4-body
propagators in Eqs. (\ref{FYE1}) and (\ref{FYE2}), which are used to
regulate the FY equations for the contact potential. The two-boson
scattering amplitude (\ref{texp}), which are introduced in the FY
equations above, carries the effective range correction in the
calculations that are presented in the next section.

The projected 4-body free resolvent, ${\cal G}^{(4)}$, are
subtracted at an energy $-\mu^2_4$:
\begin{eqnarray}
{\cal G}^{(4)}_{ij;ik}=\langle\chi_{ij}|
\left(\left[E-H_0\right]^{-1}-\left[-\mu_{4}^2- H_0\right]^{-1}
\right)|\chi_{ik}\rangle \label{G4}
\end{eqnarray}
where $H_0$ is now the free four-body Hamiltonian. Notice that when
introducing ${\cal G}^{(3)}_{ij;ik}$, eq. (\ref{G3}), in the FY
equations, the free three-body hamiltonian should be substituted by
the four-body one. The energy of the two-body subsystem $(ij)$
appearing as arguments of the two-boson scattering amplitude in
Eq.s~(\ref{FYE1}) and (\ref{FYE2}), are $\epsilon_{ij,k}^{\,\,\,l}$
and $\epsilon_{ij,kl}$, associated with a virtual pair in the 3+1
partition and in the 2+2 partition, respectively.

The subtraction scale $-\mu_{3}^2$ in ${\cal G}^{(3)}_{ij;ik}$ the
right-hand side of Eq. (\ref{FYE1}), fix the trimer properties
consistently with the STM equation, but independently on the scale
$\mu_4$. Looking closer to this term and taking into account the
left hand side, the position of the trimer pole is guarantee to be
same as the bound state energy obtained by solving the subtracted
STM equation. The other terms in the  FY equations, in (\ref{FYE1})
and (\ref{FYE2}), are regularized with an independent subtraction
scale $-\mu^2_4$. This method ensures that the three-body
subtraction scale fixes the three-body properties consistently with
the subtracted STM equation, whereas the regularization parameter
$\mu_4$ fixes the four-body observables~\cite{YamEPL06}.

\section{Position of four-body resonant loss peaks\label{sec:3}} 
The four-atom
recombination resonates when the tetramer has zero energy and placed
at the four-body continuum threshold. In cold atom traps the
resonance is reached by tuning the negative scattering length close
to a Feshbach resonance (see e.g. \cite{FerPHYS2010}). The
theoretical description of the  scaling function, which correlates
the values of the negative scattering lengths where successive
tetramers reach the continuum, introduced in \cite{hadi2011}, is now
extended to include the effective range, as
\begin{eqnarray}
a^T_{N_3,N+1}=a^-_{N_3} {\cal A}
\left({\frac{a^T_{N_3,N}}{a^-_{N_3}}},{\frac{r_0}{a^-_{N_3}}}\right)
\ , \label{eqa}
\end{eqnarray}
where $a^-_{N_3}$ is the position of the peak of the three-atom
resonant recombination for $a<0$, and $a^T_{N_3,N}$ is the
scattering length for which the excited $N$-th tetramer dissociates
and meets the four-body continuum.
The solutions of the subtracted STM (\ref{FE1}) and FY
equations (\ref{FYE1},\ref{FYE2}) depend only on $a$, $r_0$, and on
the three- and four-body scale parameters. Then for the zero energy
tetramer, the value of the negative scattering length depends only
on $r_0$ and the short-range three- and four-body momentum scales at
the subtraction points. The three-body scale can be parameterized by
$a^-_{N_3}$, while the four-body one by $a^T_{N_3,N}$ in the
determination of $a^T_{N_3,N+1}$. In this way (\ref{eqa}) can be
built. Furthermore, the scaling function $\cal A$ is calculated with
the two-boson scattering amplitude (\ref{texp}) and the solutions of
the FY equations, as well as of the STM equation, expanded up to
first order in $r_0$. 

The dimensionless function $\cal A$ is determined by solving the
subtracted STM \cite{AdhPRL95} and FY \cite{YamEPL06} equations
considering the expansion of the atom-atom $s-$wave amplitude up to
order $k^2$ as given by (\ref{texp}) and the corresponding relation
between the scattering length and virtual state energy (\ref{a}).
The solutions of the subtracted FY equations depend only on $a$,
$r_0$, and on the three- and four-body scale parameters. Then, for
the zero energy tetramer, the value of the negative scattering
length depends only on $r_0$ and the short-range three- and
four-body momentum scales at the subtraction points. The three-body
scale can be parametrized by $a^-_{N_3}$, when solving the STM
equation with $r_0$ included, while the four-body scale is
correlated to $a^T_{N_3,N}$ in the determination of $a^T_{N_3,N+1}$.
Then, the scaling function (\ref{eqa}) is built without reference to
the scale ratio, by eliminating this dependence in the ratio
$a^T_{N_3,2}/a^-_{N_3}$ in terms of $a^T_{N_3,1}/a^-_{N_3}$.

\begin{figure}[thb]
\vspace{-1cm}
\hspace{-1cm}\includegraphics[width=3.5in]{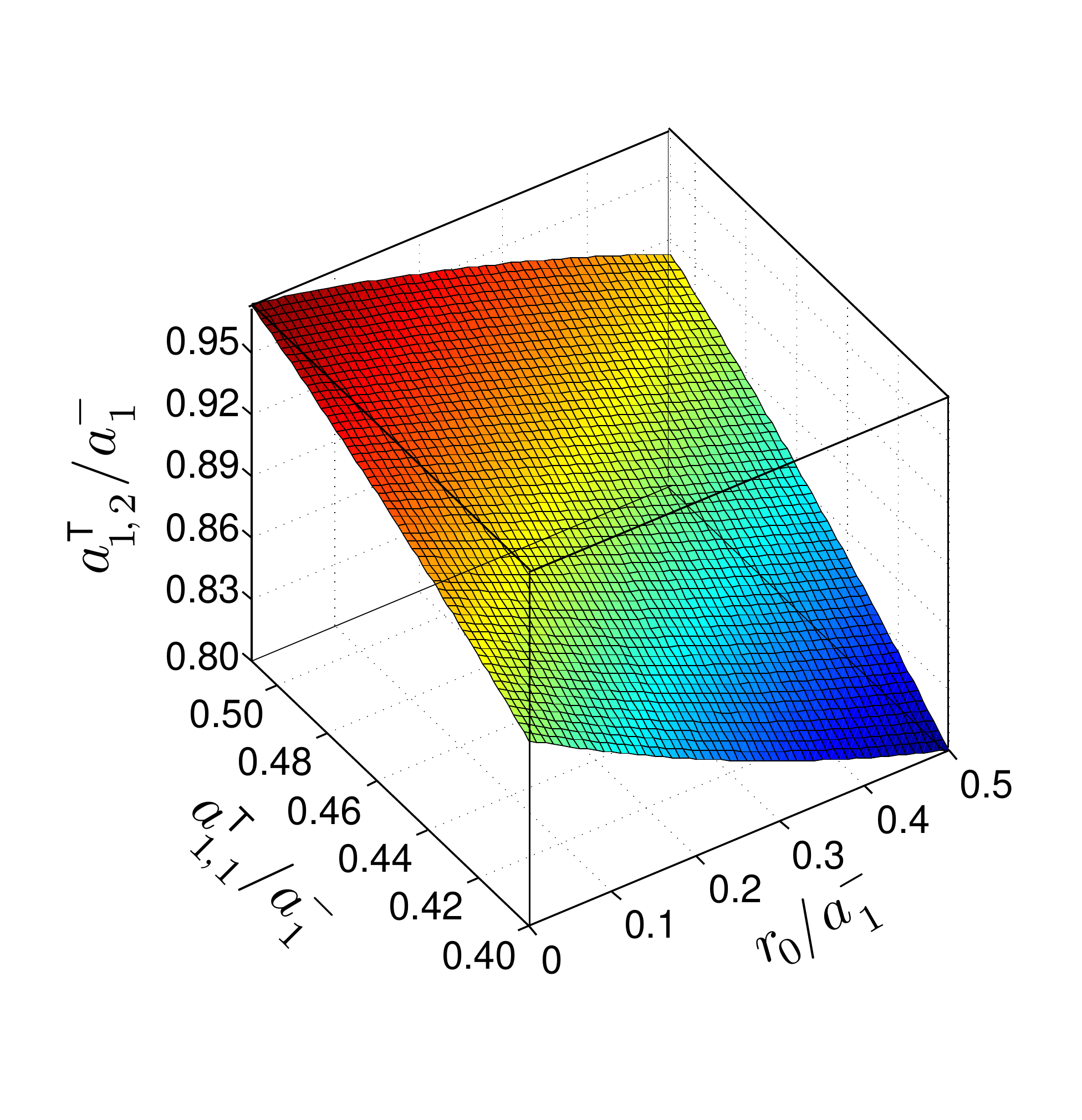}
\vspace{-1cm} \caption{{\it (color online)} The scaling function
$\cal A$ expressing the dependence of $a^T_{1,2}/a_1^-$ in
$a^T_{1,1}/a_1^-$ and $r_0/a_1^-$.  Calculations are performed in a
leading order expansion of the solutions of the FY equations with
the two-body amplitude (\ref{texp}), see text for further
explanation. } \label{fig1}
\end{figure}

In our calculations we have only obtained results for tetramers
below the ground state trimer, which should be enough for our study,
as the values of the scattering length are much larger than the
short-range length scales, corresponding to the three- and four-body
subtraction points. The scaling function ${\cal
A}\left(a^T_{1,1}/a^-_1,r_0/a^-_1\right)$, which provides the
correction to the position of the four-atom resonance is shown in
the 3D plot of Fig.~\ref{fig1}. We have presented results up to
$r_0/a^-_1\sim 0.5$, for 0.40$\le a^T_{1,1}/a^-_1\le 0.51$,
considering the relevant region where recombination data for cesium
have been measured. In this case, $r_0\sim a^T_{1,1}$, and for the
validity of the expansion in  the effective range one should
have $r_0<< a^T_{1,1}$, and the expansion of the scattering length
(\ref{a}) is questionable. Fortunately, we found that the
coefficients for the  effective range correction are fairly
small, allowing to extend the validity of the expansion to the
region where data are found.

Within the region of the plot given in Fig.~\ref{fig1}, the surface
can be parameterized by
\begin{equation}
\frac{a^T_{1,2}}{a^-_1}={\cal A}(x,y)=\sum_{0\le m+n\le2}c_{mn}\, (x-0.45)^m
y^{n}, \label{parameter}
\end{equation}
where $x\equiv a^T_{1,1}/a^-_1$,  $y\equiv r_0/|a^-_1|$, and
$(m,n)\ge 0$. The coefficients are given in the following
Table~\ref{table1}.

\begin{table}[hbt]
\caption {Coefficients for the parametrization (\ref{parameter}).
} \label{table1}
\begin{tabular}{cccccc}
\hline \hline
$c_{00}$ & $c_{10}$ & $c_{01}$ & $c_{11}$ & $c_{20}$ & $c_{02}$
\\ \hline\hline
    0.932    &   0.724   & -0.144 &  0.347 & -0.645  & 0.001
\\ \hline \hline
  \end{tabular}
\end{table}

Although the expansion (\ref{parameter}) contains
nonlinear terms in $r_0$, it is built in linear order in the
solutions of the subtracted STM and FY equations. The second order
terms are induced by the curvature of ${\cal A}(x,y)$ in $x$ and
$y$. In our case the calculation is performed for the ground state
trimer and two successive tetramer states.

The coefficients for the  effective range correction are
expected to be smaller than the ones associated with the variation
of $a^T_{1,1}/a^-_1$. Indeed, the expansion around
$a^T_{1,1}/a^-_1=0.45$ shows that the linear coefficient for
$r_0/a^-_1$ correction gives $c_{01}/c_{00}=-0.15$, while
$c_{10}/c_{00}=0.78$ is 5 times larger. The amazing smallness of
$c_{01}$ and $c_{02}$ with respect to the other coefficients,
reflects that a fraction of the effect from the  effective
range is absorbed by the variation of the short-range four-body
scale, but not all. Given that, it is quite obvious that the
correction coefficients for the  effective range contribution,
shown in Table~\ref{table1}, should be somewhat smaller than the
coefficients associated with the expansion parameter
$a^T_{1,1}/a^-_1$.

The negative value of $c_{01}$ decreases the function $\cal A$ for
positive effective ranges, which  suggests that the region where the
two-body amplitude (\ref{texp}) increases with respect to the
zero-range one, for $a<0$ and positive  effective ranges, is
important for the tetramers.  In this case, larger variations of the
scattering length toward negative values, compared to the trimer
value of $a^-_1$, are required to tune the tetramers to zero energy.

\begin{figure}[hbt]
\begin{center}
\includegraphics[width=3.4in]{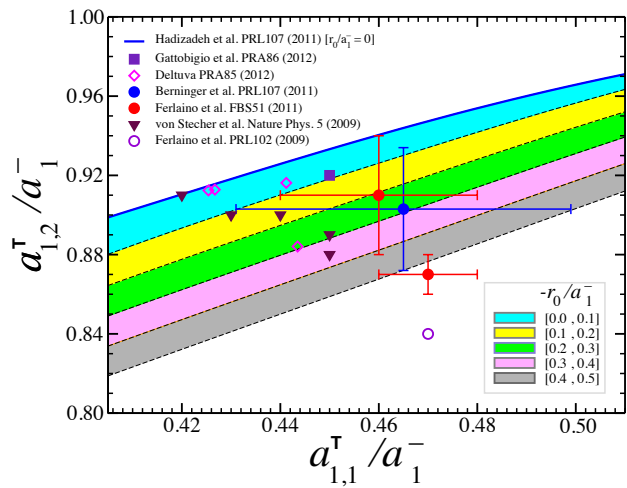}
\end{center}
\caption{{\it (color online)} Lower-order effective range
correction for the scaling plot $a^T_{1,2}/a^-_1={\cal
A}(a^T_{1,1}/a^-_1\, , \, r_0/a^-_1)$ for the correlation between
the positions of the peaks for successive tetramer resonances at the
continuum threshold, compared to other theoretical results and
experimental data. We present our results for different $r_0$
intervals (the upper solid line is from \cite{hadi2011}) compared to
a few recent calculations as indicated inside the frame
[Refs.~\cite{stechernature} (triangles), \cite{DelPRA12} (diamonds)
and \cite{GatArX12} (square)]. The experimental data of
Refs.~\cite{ferlaino2009}, \cite{FerFBS11} and \cite{BerPRL11} are
also indicated inside the figure. The lowest band (gray) corresponds
to $r_0$ for the van der Waals like potential for cesium atoms with
values within the interval $0.4\,\lesssim - r_0/a^-_1\lesssim\,0.5$
for a reference value of $a^-_1\simeq-9\,R_{\mbox{vdW}}$ (see Table
\ref{table2}).} \label{fig2}
\end{figure}

Two theoretical results are used for comparison with the results of
the scaling function $\cal A$, using the plot of Fig.~(\ref{fig2}),
the first one by Deltuva~\cite{DelPRA12} (0.4435,0.8841), which
quotes $r_0/|a^-_1|=0.33$~\cite{DelPriv} compared to 0.29  extracted
from the plot, and another one by von Stecher~\cite{stechernature},
(0.45,0.88), with $r_0/|a^-_1|=$0.38~\cite{SetPriv} compared to
0.36 obtained from
the plot. This comparison with completely different models,
separable potential model \cite{DelPRA12} and local Gaussian
potential~\cite{stechernature}, gives us confidence on the
universality and utility of the function $\cal A$, to analyze the
data. This comparison suggests a lower bound of $|r_0/a^-_1|> 0.02$
in our extraction method.

In Fig.~\ref{fig2}, the 2D plot of our results for the surface
${\cal A}(x,y)$, corresponding to the scaling function (\ref{eqa}),
is shown. The bands in the plot gives the shift of the scaling
function due to the finite values of $r_0/|a^-_{N_3}|$ in the
intervals depicted in the figure. For our reference we also plot the
theoretical results from \cite{stechernature} and \cite{DelPRA12},
and the experimental values from Refs.
\cite{ferlaino2009,FerFBS11,BerPRL11}. The data from Ref.
\cite{pollack} have large error bars and suggests $r_0$ ranging from
negative to positive values.

\section{Analysis of the cesium data\label{sec:4}} The effective range extracted
from the experimental data is shown in Table~\ref{table2}. They were
obtained just by inspection of Fig.~\ref{fig2}, by comparing the
experimental values with the scaling curve including the 
effective range correction. The shift of the data yields values of
$r_0/a^-_1$ larger than the lower bound for the theoretical
extraction. We didn't present results for \cite{pollack} as it has
larger errors. We just notice that the effective ranges from these
data varies from negative to positive values. The extraction of
$r_0$ from the data for cesium atoms close to Feshbach resonances
are within errors consistent. The extracted values of $r_0$  vary
from 2 to 5 times the van der Waals length ($R_{\mbox{vdW}}$), which
for the $Cs_2$ system is $R^{Cs_2}_{\mbox{vdW}}=\,101.0\,a_0$
\cite{ChinRMP10}.

The effective range of potentials with van der Waals tail
$-C_{6}/r^{6}$ at large distances has the approximate formula given
by ~\cite{GribPRA93,GaoJPB04,ChinPRA04}:
\begin{equation}
r_{0}=\frac{2}{3} \frac{\Gamma(1/4)^{4}}{(2\pi)^{2}}\,{\bar{a}} \,\,
\vartheta \,\, {\text{and}}\,\,
\vartheta=\left(\frac{\bar{a}}{a}\right)^{2} +
\left(\frac{\bar{a}}{a}-1\right)^{2} \ , \label{eq:BoGaoFormula}
\end{equation}
where $\bar{a}=2\pi/\Gamma(1/4)^{2}\,R_{\mbox{vdW}}\simeq0.955978
\,R_{\mbox{vdW}} $ is the average scattering length of van der Waals
potentials~\cite{ChinPRA04}, with the van der Waals length
$R_{\mbox{vdW}}=\frac12(mC_{6}/\hbar)^{1/4}$. Considering the van
der Waals length one gets that $\bar{a}= 96.5\, a_0$, and
\begin{equation}
r_{0}\simeq
2.7894\,R_{\mbox{vdW}}\,\vartheta=2.9179\,\bar{a}\,\vartheta \simeq
281.7 \,\vartheta\, a_0 . \label{eq:BoGaoFormula-1}
\end{equation}
In the actual cold cesium experiments listed in Table \ref{table2},
the factor $\vartheta$ varies from  1.26 (for the second tetramer
resonance) to 1.55 (for the first tetramer resonance), which gives
$r_0$ within $3.5\,-\,4.3\,R_{\mbox{vdW}}$. The corresponding
interval of $0.4\,\lesssim r_0/|a^-_1|\lesssim\,0.5$ for a reference
value of $a^-_1\simeq-9\,R_{\mbox{vdW}}$, can be identified with the
lowest band (gray) shown in Fig.~\ref{fig2}.

The estimated values of $r_0$ extracted from the shift of the data
with respect to the zero-range calculations, shown in
Fig.~\ref{fig2}, are given in Table~\ref{table2}. The values for the
effective range are found between 2$\,R_{\mbox{vdW}}$ and
$\sim\,5\,R_{\mbox{vdW}}$, while Eq. (\ref{eq:BoGaoFormula-1})
applied to the set of cesium data suggests the interval
$3.5\,-\,4.3\,R_{\mbox{vdW}}$ represented by the lowest (gray) band
in the figure. The present experimental errors overlaps the cesium
data, as one verify in Fig.~\ref{fig2}. By performing the error
weighted average of the extracted effective range values from the
data given in Refs.~\cite{BerPRL11} and \cite{FerFBS11} we obtain
3.9$\pm 0.8\,R_{\mbox{vdW}}$. In particular, the data set
$\{0.47(1), 0.87(1)\}$, provides an effective range of about
4.8$\,R_{\mbox{vdW}}$, which within $1\sigma$ is consistent with the
values of the van der Waals potential. The data set
$\{0.47,0.84\}$\cite{ferlaino2009} with no quoted errors suggests a
large value of $r_0>5\,R_{\mbox{vdW}}$.

\begin{table}[hbt]
\caption {Extracted effective ranges from the shift of the
experimental peaks of the four-atom losses for a gas of cold cesium
atoms close to a Feshbach resonance, using the scaling plot shown in
Fig.~\ref{fig2}. The errors in the extracted values of $r_0$ are
estimated from the figure as well.} \label{table2}
\begin{tabular}{ccccc}
\hline \hline Ref. & $a_{1,1}^{T}/a_{1}^{-} $ &
$a_{1,2}^{T}/a_{1}^{-} $ & $a_{1}^{-}$ [$R_{\mbox{vdW}}$] & $r_0$
[$R_{\mbox{vdW}}$] \\
\hline\hline
\cite{ferlaino2009} & 0.47  & 0.84 &     -8.7(1)    &    $>5$ \\
\cite{BerPRL11}  & 0.465(34)    & 0.903(31) &       -9.54(28)   &  $2.5\,\pm\,1.7$  \\
\cite{FerFBS11}  & 0.47(1)  &  0.87(1) &    -8.71    & $4.8 \,\pm\,1.0$   \\
\cite{FerFBS11}  & 0.46(2)  &  0.91(3) &    -9.64    & $2\,\pm\,2$
\\ \hline \hline
\end{tabular}
\end{table}

It is noticeable that the dissociation point of cesium ground state
Efimov trimers for negative scattering lengths are found in a narrow
band\cite{BerPRL11}, independently on which Feshbach resonance the
system is tuned (see Table \ref{table2}). Theoretical
works~\cite{WanPRL12,NaiPRA12,NaiArxiv12,SchArxiv12,SorPRA12}
addressed the interesting issue of the physical mechanism for the
dominance of the two-body properties on the position of the first
Efimov resonance for $a<0$, when the trimer meets the continuum. The
results in Table \ref{table2} show a range of values for $r_0$
extracted from the tetramer and trimer resonances, which within
errors (excepting one data set) are consistent with the effective
range values from the van der Waals type potential. The numerical
analysis of Wang et al. \cite{WanPRL12} has shown that two-body
interactions, which suppress efficiently the wave function for
separation distances less than $r_0$,  have
$a^-_1\sim-9\,R_{\mbox{vdW}}$, determined by two-body properties,
with this class of systems closely related to entrance-channel
dominated Feshbach resonances. The dominance of the single channel
potential with the van der Waals tail in the three-cesium reactions,
with suppression of the short distance wave function, putted forward
the narrow band where the trimer dissociation position is found. It
also suggests that the effective range seen through the shift of the
position of the four-atom recombination peak with respect to the
zero-range results, should be given by the van der Waals tail of the
potential, as well.

It is worthwhile to address the validity of  the correction
due the effective range, considering the main results presented in
Fig.~\ref{fig2} and Table \ref{table2}. In view of the expansion
given in Eq.(\ref{texp}), which at a first glance would be doubtful,
when $r_0/|a^-_1| \sim a^T_{12}/a^-_1$, we found consistence between
the present calculations with the results obtained with short-range
separable\cite{DelPriv} and local\cite{SetPriv} interactions. The
leading coefficient for the expansion in $r_0/|a^-_1|$ is one fifth
of the corresponding coefficient for the dependence of the scaling
function in $a^T_{11}/a^-_1$ (see Table \ref{table1}), which shows
that the extracted values of $r_0$ using the scaling plot comes from
small corrections to the zero-range scaling function. Therefore, as
observed in Fig.~\ref{fig2}, the agreement of the expansion we have
considered can be extended also to cases near $a^T_{12}/a^-_1\sim
r_0/|a^-_1|$.

\section{Conclusion and Outlook\label{concl}}
In this work, we present an approach to extract the effective
range of a given atomic system from the shifts in the four-body
recombination peaks, at the tetramer dissociation threshold, with
respect to the zero-range results. We solved the trimer and tetramer
subtracted zero-range integral equations with two-body amplitudes
carrying the effective range in lowest order. Besides, the trimer
and tetramer subtracting scales, no further parameters are
introduced in the calculations. The correlation between the negative
scattering lengths where successive tetramers dissociates at the
four-body continuum is extended to include the correction due to the
effective range, which is model independent checked against
different short-range potential model calculations. We found that
the  effective range correction of the on-shell two-body
amplitude in the calculation of the trimer and tetramer dissociation
points, when presented as a correlation plot of the successive
positions of the tetramer resonances is not completely parameterized
by the trimer and tetramer short-range scales, as one could naively
expect. This unexpected property turned to be essential to
 single out the effective range from the shift in the positions
of the four-atom recombination peaks with respect to the zero-range
theory.

We applied our proposal to the cesium data measured at broad
$s-$wave Feshbach resonances~\cite{ferlaino2009,BerPRL11,FerFBS11}.
In our analysis we considered the shifts of the observed
recombination peaks at the tetramer dissociation positions with
respect to the zero-range results. The effective ranges were found
within the interval $2\,R_{\mbox{vdW}}\, \lesssim\, r_0\, \lesssim
\, 5\,R_{\mbox{vdW}}$, with a reference value of 3.9$\pm 0.8\,
R_{\mbox{vdW}}$, obtained by the error weighted average of the
$r_0$'s extracted from the data of Refs.~\cite{BerPRL11,FerFBS11}.
These values are consistent with the van der Waals potential tail
for the broad $s-$wave Feshbach resonances of the $Cs_2$ system,
providing $r_0$ within the interval of $3.5\,-\,4.3\,R_{\mbox{vdW}}$
for the scattering lengths where the resonances were found. Our
results put a strong evidence on the prevalence of the
entrance-channel dominance in the physics of the universal tetramers
formed with broad Feshbach resonances.

The question on how the effective range is sensible to Feshbach
resonance manipulations in the scattering length as suggested by
(\ref{eq:BoGaoFormula-1}) has to be considered. Once improved
experimental data are available, it will be possible to verify in
more detail how Feshbach resonances from different setups affect the
$r_0$ values extracted from the trimer and tetramer dissociation
points. The present analysis, based on the universality of 
effective range correction to the position of tetramer resonances,
can be generally applied to any other cold-atom bosonic system close
to Feshbach resonances.

\acknowledgments
We thank A. Deltuva and J. von Stecher for sending us some of their
effective range results. TF and LT also appreciate clarifying
discussion with P. Naidon and S. Endo. Our thanks also to Funda\c
c\~ao de Amparo a Pesquisa do Estado de S\~ao Paulo and Conselho
Nacional de Desenvolvimento Cient\'\i fico e Tecnol\'ogico for
partial support.

\end{document}